\documentclass[a4paper,12pt]{article}
\usepackage{epsfig}
\newcommand{\bmat}{\left(\begin{array}}
\newcommand{\emat}{\end{array}\right)}
\def\NPB#1#2#3{Nucl. Phys. B{#1} (19#2) #3}
\def\PLB#1#2#3{Phys. Lett. B{#1} (19#2) #3}

\def\PRD#1#2#3{Phys. Rev. D{#1} (19#2) #3}
\def\PRL#1#2#3{Phys. Rev. Lett. {#1} (19#2) #3}

\def\lsim{\raise0.3ex\hbox{$\;<$\kern-0.75em\raise-1.1ex\hbox{$\sim\;$}}}
\def\gsim{\raise0.3ex\hbox{$\;>$\kern-0.75em\raise-1.1ex\hbox{$\sim\;$}}}

\def\yzero{\smash{\hbox{$y\kern-4pt\raise1pt\hbox{${}^\circ$}$}}}

\def\-{\hphantom{-}}

\def\s2{\frac{1}{\sqrt2}}

\def\beq{\begin{equation}}
\def\eeq{\end{equation}}
\def\beqa{\begin{eqnarray}}
\def\eeqa{\end{eqnarray}}

\def\IF{\relax{\rm I\kern-.18em F}}
\def\II{\relax{\rm I\kern-.18em I}}
\def\IP{\relax{\rm I\kern-.18em P}}
\def\IC{\relax\hbox{\kern.25em$\inbar\kern-.3em{\rm C}$}}
\def\IR{\relax{\rm I\kern-.18em R}}

\def\Dsl{\,\raise.15ex\hbox{/}\mkern-13.5mu D} 
\def\IZ{Z\kern-.4em  Z}
\def\bmat{\left(\begin{array}}
\def\emat{\end{array}\right)}

\def    \part          {\partial}
\def    \be            {\begin{equation}}
\def    \ee            {\end{equation}}
\def    \bea           {\begin{eqnarray}}
\def    \eea           {\end{eqnarray}}

\begin{document}
%
\pagestyle{empty}
\rightline{FTUAM 00/12}
\rightline{IFT-UAM/CSIC-00-21}
\rightline{hep-ph/0006266}
\rightline{June 2000}

\renewcommand{\thefootnote}{\fnsymbol{footnote}}
\setcounter{footnote}{0}

\vspace{0.0cm}
\begin{center}
\large{\bf Initial Scales, Supersymmetric Dark Matter and
Variations of Neutralino-Nucleon Cross Sections\\[5mm]}

\large{E. Gabrielli$^{1,3}$\footnote{emidio.gabrielli@cern.ch}, S.
Khalil$^{1,2}$\footnote{shaaban.khalil@uam.es},  
C. Mu\~noz$^{1,3}$\footnote{carlos.munnoz@uam.es},
E. Torrente-Lujan$^{1}$\footnote{e.torrente@cern.ch} 
}

\begin{center}
{\small
{\it $^1$ Departamento de F\'{\i}sica
Te\'orica C-XI, Universidad Aut\'onoma de Madrid,\\[-0.1cm]
Cantoblanco, 28049 Madrid, Spain. \\
\vspace*{2mm}
\it $^2$ Ain Shams University, Faculty of Science, Cairo
11566, Egypt.} \\[2mm]
{\it $^3$
Instituto de F\'{\i}sica Te\'orica  C-XVI,
Universidad Aut\'onoma de Madrid,\\[-0.3cm]
Cantoblanco, 28049 Madrid, Spain.} 
}
\end{center}

{\bf Abstract} 
\\[7mm]
\end{center}
\begin{center}
\begin{minipage}[h]{14.0cm}
The neutralino-nucleon cross section in the context of the
MSSM with universal soft supersymmetry-breaking terms 
is compared with the limits from dark matter detectors. 
Our analysis is focussed on the stability of the corresponding
cross sections with respect to variations of the initial scale
for the running of the soft terms, finding
that the smaller the scale is, the larger the cross sections become.
For example, by taking
$10^{10-12}$ GeV rather than 
$M_{GUT}$, which is a more 
sensible choice, in particular in the context of some superstring models,
we find 
extensive regions in the parameter space 
with cross sections 
in the range of $10^{-6}$--$10^{-5}$ pb, i.e. where current dark matter
experiments are sensitive.
For instance, this can be obtained for $\tan\beta\gsim 3$.

\end{minipage}
\end{center}
\vspace{1.0cm}
\begin{center}
\begin{minipage}[h]{14.0cm}

\end{minipage}
\end{center}
\newpage
\setcounter{page}{1}
\pagestyle{plain}
\renewcommand{\thefootnote}{\arabic{footnote}}
\setcounter{footnote}{0}
%
%
\section{Introduction}

Recently 
there has been some theoretical activity 
\cite{Bottino,arna,Ellis,arnowitt,nath,focus} 
\footnote{See also \cite{Drees,antiguos,Baer,cuerdas} for other works.} 
analyzing the compatibility of regions
in the parameter space of 
the minimal supersymmetric standard model (MSSM)
with the sensitivity of current dark matter detectors,
DAMA \cite{experimento1} and CDMS \cite{experimento2}.
These detectors are sensitive to a 
neutralino-nucleon cross 
section\footnote{Let us recall that the lightest neutralino 
$\tilde\chi_1^0$ is the natural candidate
for dark matter in supersymmetric theories with conserved R parity,
since it is usually the lightest supersymmetric particle (LSP) and
therefore stable \cite{review}.
}
$\sigma_{\tilde\chi_1^0-p}$ in the range of $10^{-6}$--$10^{-5}$ pb.
Working in the supergravity framework for 
the MSSM with universal soft terms, 
it was pointed out 
in \cite{Bottino,arna,arnowitt} 
that the large $\tan\beta$ regime   
allows regions where the above mentioned range 
of $\sigma_{\tilde\chi_1^0-p}$ is reached.
Besides, working with non-universal soft scalar masses,
they also found $\sigma_{\tilde\chi_1^0-p}\approx 10^{-6}$
pb for small values of $\tan\beta$.
In particular, this was obtained for $\tan\beta\gsim 25$ 
($\tan\beta\gsim 4$) working with universal (non-universal) soft terms
in \cite{arnowitt}.
The case of non-universal gaugino masses was also analyzed in \cite{nath} 
with interesting results.

The above analyses  
were performed assuming universality (and
non-universality)
of the soft breaking terms at the unification scale, 
$M_{GUT}\approx 10^{16}$ GeV, as it is usually done in the
MSSM literature. 
Such a scale can be obtained in a natural manner within superstring
theories. This is e.g. the case of type I superstring theory 
\cite{Witten,Rigolin} 
and heterotic M-theory \cite{Witten,Banks}.
However, recently, going away from perturbative
vacua, it was realized that the string scale may be anywhere between
the weak scale and the Planck scale. 
For instance D-brane configurations where the standard model lives,
allow these possibilities in type I 
strings \cite{Lykken,Dimopoulos,typeITeV,typeIinter}.
Similar results can also be obtained in type II strings \cite{typeII}
and weakly and strongly coupled heterotic strings 
\cite{stronghete,weakandstronghete}.
Hence, it is natural to wonder how much the standard neutralino-nucleon
cross section analysis will get modified by taking 
a scale $M_I$ smaller than $M_{GUT}$
for the initial scale of the soft terms \footnote{This question was
recently pointed out in \cite{Allanach} for a different type of
analysis.
In particular the authors studied low-energy implications, 
like sparticle spectra and charge
and colour breaking constraints, of
a string theory with a scale of order $10^{11}$ GeV. Similar phenomenological 
analyses were carried out
in the past \cite{Pomarol} for $M_{Planck}$ rather than $M_{GUT}$.}.

The content of the article is as follows. In Section 2
we will briefly review 
several scenarios suggested by superstring theory
where the initial scale 
for the running of the soft terms is $M_I$ instead of $M_{GUT}$. 
In particular, we will see that $M_I\approx 10^{10-14}$ GeV
is an attractive possibility. The issue of gauge coupling unification,
which is important for our calculation, will also be discussed.
Then, in Section 3, we will study in detail
the stability of the neutralino-nucleon cross section with
respect to variations of $M_I$.  
For the sake of generality,
we will allow $M_I$ to vary
between $10^{16}$ GeV, which corresponds to $M_{GUT}$, and $10^{10}$ GeV.
Of course the results will be valid not only for low-scale string scenarios
but also for any scenario 
with an unification scale smaller than $M_{GUT}$.
Let us finally remark that 
the analysis will be carried out for the  
case of universal soft terms. 
This is the most simple situation in the framework of the MSSM
and can be obtained e.g. in superstring models with 
dilaton-dominated supersymmetry breaking \cite{dilaton} or in weakly
and
strongly coupled heterotic
models with one K\"ahler modulus \cite{softM}.
Finally, the conclusions are left  
for Section 4.

\section{Initial scales}

As mentioned in the Introduction, it was recently realized 
that the string scale is not necessarily close to the Planck
scale but can be as low as the electroweak scale.
In this context, two 
scenarios are specially attractive in order to attack the
hierarchy problem of unified theories: a non-supersymmetric  
scenario with the string scale of order a few TeV 
\cite{Dimopoulos,typeITeV},
and a supersymmetric scenario with the string scale of order 
$10^{10-12}$ GeV \cite{typeIinter}.
Since we are interested in the analysis of supersymmetric dark matter, 
we will concentrate on the latter.

In supergravity models supersymmetry can be spontaneously broken in a 
hidden sector of the theory and the gravitino mass,
which sets the overall scale of the soft terms, is given by:
\bea
m_{3/2}\approx \frac{F}{M_{Planck}}\ ,
\label{gravitino}
\eea
where $F$ is the auxiliary field whose vacuum expectation value
breaks supersymmetry. 
Since in supergravity one would expect $F\approx M_{Planck}^2$, one  
obtains
$m_{3/2}\approx M_{Planck}$ and therefore 
the hierarchy problem solved in principle by supersymmetry
would be re-introduced,
unless non-perturbative effects such as gaugino condensation
produce $F\approx M_W M_{Planck}$. However, if the scale
of the fundamental theory is $M_I\approx 10^{10-12}$ GeV instead of
$M_{Planck}$,
then $F\approx M_I^2$
and one gets $m_{3/2}\approx M_W$ in a natural way, without invoking any
hierarchically suppressed non-perturbative effect \cite{typeIinter}.

For example, embedding the standard model inside D3-branes in type I
strings,the string scale is given by:
\bea
M_I^4= \frac{\alpha M_{Planck}}{\sqrt 2} M_c^3\ ,
\label{gravitino2}
\eea
where $\alpha$ is the gauge coupling and $M_c$ is the compactification scale. 
Thus one gets $M_I\approx 10^{10-12}$ GeV with $M_c\approx 10^{8-10}$ GeV.

There are other arguments in favour 
of scenarios with initial scales $M_I$ smaller than $M_{GUT}$.
For example in \cite{stronghete} scales $M_I\approx 10^{10-14}$ GeV
were suggested to explain many experimental observations as
neutrino masses or the scale for axion physics. These scales might also
explain the observed ultra-high energy ($\approx 10^{20}$ eV) cosmic rays
as products of long-lived massive string mode decays. Besides,
several models of chaotic inflation favour also these 
initial scales \cite{caos}.

Inspired by these scenarios
we will allow the initial scale $M_I$ for the running of the soft terms
to vary between $10^{16}$ GeV and $10^{10}$ GeV,
when computing the neutralino-nucleon cross section below.
As we will see, the values of the gauge coupling constants at those
scales will be crucial in the computation.
This head us for a brief discussion of gauge coupling unification in models
with low initial scale:

(a) {\it Non universality of gauge couplings}

\noindent An interesting proposal  
in the context of
type I string models was studied in \cite{Rigolin,mirage}: 
if the standard model comes from the
same collection of D-branes, 
stringy corrections might change the boundary conditions at the string scale
$M_I$ to mimic the effect of field theoretical logarithmic running.
Thus the gauge couplings will be non universal and their values
will depend on the initial scale $M_I$ chosen. 
This is schematically shown in Fig.~1a for the scale $M_I=10^{11}$ GeV,
where $g_3\approx 0.8$, $g_2\approx 0.6$ and $g_1\approx 0.5$.
Clearly, another possibility giving rise to a similar result
might arise when
the gauge groups came from different types of D-branes.
Since different D-branes have associated different couplings,
this would imply the non universality of the gauge couplings. 

(b) {\it Universality of gauge couplings}

\noindent On the other hand, if gauge coupling 
unification at $M_I$, $\alpha_i=\alpha$, is what we want to obtain, then 
the addition of extra fields
in the massless spectrum can achieve this task \cite{typeIinter}.
An example of additional particles which can produce the
beta functions, $b_3=-3$, $b_2=3$, $b_1=19$, 
yielding unification at around $M_I=10^{11}$ GeV
was given in \cite{Allanach}
%
\bea
2\times [(1,2,1/2)+(1,2,-1/2)]
+ 3\times [(1,1,1)+(1,1,-1)]\ ,
\label{example11}
\eea
where the fields transform under $SU(3)_c\times SU(2)_L\times U(1)_Y$.
In this example one obtains $g(M_I)\approx 0.8$. This is schematically
shown in Fig.~1b.

As a matter of fact, once the Pandora's box containing extra
matter fields is open, 
other possibilities arise. Note that the example above does not contain
extra triplets, and therefore $\alpha_3$ runs as in the MSSM
\bea
\frac{1}{\alpha_3 (Q)}=\frac{1}{\alpha_3 (M_{susy})} - \frac{b_3}{2\pi}\ln
\frac{Q}{M_{susy}}\ ,
\label{example0}
\eea
where $b_3=-3$ and $M_{susy}$ indicates the supersymmetric threshold.
However, introducing $n_3$ extra triplets,
$b_3=-3+\frac{1}{2}n_3$ will increase and therefore $\alpha_3$ will also
increase.
Likewise, extra doublets and/or singlets will allow to increase the
value of $\alpha_2$ and $\alpha_1$ as in the example above and therefore
we will be able to obtain  
unification at $M_I$, but for bigger values of $\alpha_i=\alpha$.
For example, the additional particles
\bea
3\times [(3,1,2/3)+(\bar 3,1,-2/3)]
+ 6\times [(1,2,1/2)+(1,2,-1/2)]\ ,
\label{example1}
\eea
produce $b_3=0$, $b_2=7$, $b_1=27$, yielding unification again at around
$M_I=10^{11}$ GeV but for $g(M_I)\approx 1.3$.

We will see in the next section that due to the different values of
the gauge couplings at $M_I$, scenarios (a) and (b) give rise 
to qualitatively different results for cross sections.

\section{Neutralino-nucleon cross sections versus initial scales}

In this section we will consider the whole
parameter space of the MSSM with the only assumption of
universality. In particular, the requirement of correct electroweak
breaking leave us with four independent parameters (modulo the
sign of the Higgs mixing parameter $\mu$ which appears in the
superpotential $W=\mu H_1H_2$).
These may be chosen as follows: $m$, $M$, $A$ and $\tan\beta$, i.e.
the scalar and gaugino masses, the coefficient of the trilinear terms,
and the ratio of Higgs vacuum expectation values $\frac{<H_2>}{<H_1>}$.

On the other hand, we will work with the usual formulas for 
the elastic scattering
of relic LSPs on protons and neutrons that can be found in the 
literature \cite{Drees,cross,Baer,Ellis}.
In particular, we will follow the re-evaluation of the 
rates 
carried out in \cite{Ellis}, using their 
central values for the hadronic matrix elements.

As mentioned in the introduction, the initial boundary conditions for
the running MSSM soft terms are usually understood at a scale
$M_{GUT}$.
Smaller initial scales, as for example 
$M_I\approx 10^{11}$ GeV, will imply larger neutralino-nucleon
cross sections. Although we will enter in more details later on, basically
this can be understood from the variation in the
value of $\mu^2$ with $M_I$, since the cross sections are very
sensitive to this value. 

Let us discuss then first the variation of 
$\mu^2$ with $M_I$.
Recalling that this value is determined by 
the electroweak breaking conditions as
\bea
\mu^2 &=&\frac{m_{H_1}^2-m_{H_2}^2\tan^2\beta}{\tan^2\beta-1}
-\frac{1}{2}M_Z^2\ ,
\label{electroweak}
\eea
we observe that, for $\tan\beta$ fixed, the smaller the initial
scale for the running is, the smaller the numerator in the
first piece of (\ref{electroweak}) becomes. 
To understand this qualitatively, let us consider 
e.g. the evolution of $m_{H_1}^2$ (neglecting for simplicity 
the bottom and tau Yukawa couplings)
\bea
m_{H_1}^2 (t)=m^2+M^2\ G(t)\ ,
\label{evolution}
\eea
where $t=\ln(M_I^2/Q^2)$,
\bea
G(t)= \frac{3}{2} f_2(t) + \frac{1}{2} f_1(t)\ ,
\label{evolution2}
\eea
and the functions $f_i(t)$ are given by:
\bea
f_i(t)= \frac{1}{b_i}\left(1-\frac{1}{\left(1+
\frac{\alpha_i(0)}{4\pi}b_it\right)^2}\right)\ .
\label{evolution3}
\eea
For $M_I=M_{GUT}$ we recover the usual case of the MSSM: 
$t\approx 61$ for $Q\approx 1$ TeV. However, for $M_I$ smaller than
$M_{GUT}$, the value of $t$ will decrease and therefore
$f_i(t)$ will also decrease,
producing a smaller value of $G(t)$.
As a consequence $m_{H_1}^2$ at $Q\approx 1$ TeV will also be smaller.
For $m_{H_2}^2 (t)$ 
the argument is similar: the smaller the initial scale for the
running is, the less important the negative contribution $m_{H_2}^2$
to $\mu^2$ in (\ref{electroweak}) becomes.

In fact, when the initial scale is decreased, 
it is worth noticing that the values 
of the gauge couplings are modified. Therefore, this effect will also 
contribute to modify the soft Higgs mass-squared 
(see e.g. (\ref{evolution3})). Let us consider first 
the case (a) with non-universal gauge couplings at $M_I$ discussed in
the previous section (see also Fig.~1a), where
$\alpha_2 (M_I)$ and $\alpha_1 (M_I)$ are smaller than 
$\alpha(M_{GUT})$.
For instance, for
$m_{H_1}^2$ this has obvious implications: $f_i(t)$ decrease 
for $M_I$ smaller than
$M_{GUT}$, 
not only because $t$ decreases but also because 
$\alpha_i(0)$ are smaller.

The above conclusions  have important consequences
for cross sections.
As is well known, when $|\mu|$ decreases the Higgsino
components, $N_{13}$ and $N_{14}$,
of the lightest neutralino
\bea
\tilde\chi_1^0= N_{11}\tilde B^0+N_{12}\tilde W^0
+N_{13}\tilde H_1^0
+N_{14}\tilde H_2^0
\label{neutralino}
\eea
increase
and therefore the spin independent cross section also increases.
We show both facts in Figs.~2 and 3.

These figures correspond to $\mu<0$. 
Opposite values of $\mu$ imply smaller cross sections
and, moreover, well known experimental constraints as those
coming from  
the $b\rightarrow s\gamma$ process highly reduce 
the $\mu>0$ parameter space.
The other parameters are chosen as follows. 
For a given $\tan\beta$ and neutralino mass, the common 
gaugino mass $M$ is essentially fixed. 
For the common scalar mass $m$ we have taken $m=150$ GeV.
Finally, for the common coefficient of the trilinear terms $A$
we have taken $A=-M$.
This relation is particularly interesting since it arises naturally
in several string models \cite{dilaton,Rigolin}. 
In any case we have checked that the cross sections and our main
conclusions are not very sensitive to the specific values
of $A$ and $m$ in a wide range. In particular this is so for 
$\mid A/M\mid \lsim 5$ and\footnote{Let us remark however
that for $m$ in the range $50-100$ GeV the neutralino is not the
LSP in the whole parameter space. In some regions the stau is the LSP.}
50 GeV $\lsim m\lsim 250$ GeV.

We have checked that our results are consistent with present 
bounds coming from accelerators and astrophysics.
The former are
LEP and Tevatron bounds on supersymmetric masses and 
CLEO $b\to s\gamma$ branching ratio measurements.
The latter are relic neutralino density bounds and will be discussed
in some detail later on. 


In Fig.~2, for $\tan\beta=10$, 
we exhibit the gaugino-Higgsino components-squared $N_{1i}^2$ 
of the LSP as a function of its mass $m_{\tilde{\chi}_1^0}$
for two different values of the initial scale, $M_I=10^{16}$ GeV  
$\approx M_{GUT}$ and 
$M_I=10^{11}$ GeV. Clearly, the smaller the scale is, the larger 
the Higgsino components become. In particular, for 
$M_I=10^{16}$ GeV the LSP is mainly Bino since $N_{11}$ is extremely 
large. The scattering channels through Higgs exchange are
suppressed (recall that the Higgs-neutralino-neutralino couplings
are proportional to $N_{13}$ and $N_{14}$) and therefore
the cross sections are small as we will see explicitly below.
As a matter of fact, the scattering channels through squark exchange
are also suppressed by the mass of the first-family squarks.
Indeed in this limit the cross section can be approximated as
\bea
\sigma_{\tilde\chi_1^0-p}
\propto \frac{m_r^2}{4\pi}\left(\frac{g'^2\sin\theta}{m_{\tilde q}^2-
m_{\tilde{\chi}_1^0}^2}\right)^2 |N_{11}|^4
\ ,
\label{cross}
\eea
where $m_r$ is the reduced mass and $m_{\tilde q}$, $\theta$
are the mass and the mixing angle of the first-family squarks
respectively.
However for
$M_I=10^{11}$ GeV the Higgsino contributions $N_{13}$ and $N_{14}$ 
become important and even dominant for 
$m_{\tilde{\chi}_1^0}\lsim 130$ GeV 
(e.g. with $\tan\beta=3$ this is obtained for 
$m_{\tilde{\chi}_1^0}\lsim 65$ GeV).
Following the above arguments this will imply larger cross sections.
Indeed scattering channels through Higgs exchange are now important
and their contributions 
to cross sections can be schematically approximated as
\bea
\sigma_{\tilde\chi_1^0-p}
\propto \frac{m_r^2}{4\pi}
\frac{\lambda_q^2}{m_h^4}
|N_{1i} \left(g'N_{11}-g_2N_{12}\right)|^2
\ ,
\label{cross2}
\eea
where $i=3,4$, $\lambda_q$ are the quark Yukawa couplings and
$m_h$ represent the Higgs masses.
It is also worth noticing 
that, for any fixed value of $M_I$, 
the larger $\tan\beta$ is, the larger the Higgsino contributions become.
The reason being that the top(bottom) Yukawa coupling 
decreases(increases) since it is proportional
to $\frac{1}{\sin\beta}$($\frac{1}{\cos\beta}$).
This implies that the negative(positive) contribution 
$m_{H_2}^2$($m_{H_1}^2$) to $\mu^2$ is less important.
The discussion of the cases with $\tan\beta>10$ is more subtle
and will be carried out below.

The consequence of these results on the cross section is shown in Fig.~3, where
the cross section as a function
of the LSP mass $m_{\tilde{\chi}_1^0}$ is plotted
for five different values of the initial scale $M_I$.
For instance, when $m_{\tilde{\chi}_1^0}=100$ GeV, 
$\sigma_{\tilde\chi_1^0-p}$ 
for $M_I=10^{11}$ GeV is two orders of magnitude larger
than for $M_{GUT}$.
In particular, for $\tan\beta=3$, one finds 
$\sigma_{\tilde\chi_1^0-p} < 10^{-6}$ pb 
if the initial scale is $M_I=10^{16}$ GeV.
However $\sigma_{\tilde\chi_1^0-p} \gsim 10^{-6}$ GeV is possible if 
$M_I$ decreases. 
For $M_I\lsim 10^{12}$ GeV,
taking into account the experimental lower bound on the lightest
chargino
mass
$m_{\tilde\chi_1^{\pm}}=90$, 
the range 70 GeV $\lsim m_{\tilde{\chi}_1^0}\lsim$ 100 GeV is 
even consistent with the DAMA limits.
As discussed above, the larger 
$\tan\beta$ is, the larger the Higgsino contributions become, and
therefore the cross section increases.
For $\tan\beta=10$ we see in Fig.~3
that the range 60 GeV $\lsim m_{\tilde{\chi}_1^0}\lsim$ 130 GeV is now 
consistent with DAMA limits. This corresponds to $M_I\lsim 10^{14}$ GeV.

Finally, we show in Fig.~3 the case
$\tan\beta=20$. Then the above range increases
50 GeV $\lsim m_{\tilde{\chi}_1^0}\lsim$ 170 GeV, corresponding
now to  $M_I\lsim 10^{16}$ GeV.
It is worth noticing here that the value of $\mu^2$
is very stable with respect to variations
of $\tan\beta$ when this is large ($\tan\beta\gsim 10$). This is
due to the fact that $\mu^2\approx -m_{H_2}^2 - \frac{1}{2}M_Z^2$ (see
(\ref{electroweak})). Since $\sin\beta\approx 1$, the top Yukawa
coupling is stable and therefore the same conclusion is obtained for
$m_{H_2}^2$ and $\mu^2$. Thus, for a given $M_I$, 
the reason for the cross section to increase
when $\tan\beta$ increases
cannot be now the increment of the Higgsino components of the LSP.
Nevertheless there is a second effect in the cross section 
which is now the dominant one: the contribution of the down-type
quark Yukawa couplings (see (\ref{cross2})) 
which are proportional to $\frac{1}{\cos\beta}$.


In Fig.~4 we plot the neutralino-proton
cross section as a function of $\tan\beta$.
Whereas large values of $\tan\beta$  are needed in the case
$M_I=M_{GUT}$ to obtain cross sections 
in the relevant region of DAMA experiment,
the opposite situation occurs in the case $M_I=10^{11}$ GeV since smaller
values are favoured.



Let us consider now the case (b) with gauge coupling unification
at $M_I$ discussed in Section 2 (see also Fig.~1b).
The result for the cross section as a function of $m_{\tilde\chi_1^0}$ is plotted
in Fig.~5 for $\tan\beta=20$. Clearly, 
the cross section increases when $M_I$ decreases. However, 
this increment is less 
important than in the previous case.
The reason being that now
$\alpha_2 (M_I)$ and $\alpha_1 (M_I)$ are bigger than 
$\alpha(M_{GUT})$ instead of smaller.
For example this counteracts the increment of $f_i(t)$ in (\ref{evolution3})
due to the smaller value of $t$ when $M_I$ is smaller.
Due to this effect, only with $\tan\beta\gsim 20$ we obtain
regions consistent with DAMA limits.

The results for the
case with extra triplets (see e.g. (\ref{example1})) are worst
since the gauge coupling at the unification point $M_I$ is bigger than above.
The value of $\sigma_{\tilde\chi_1^0-p}$ 
at $M_I$ may be even smaller than its usual value at $M_{GUT}$.

Before concluding let us discuss briefly the effect of relic neutralino 
density bounds on cross sections.
The most robust evidence for the existence of dark matter comes from 
relatively small scales.
Lower limits inferred
from the flat rotation curves of spiral galaxies 
\cite{review,salucci} are 
$\Omega_{halo}\gsim 10\ \Omega_{vis}$ or 
$\Omega_{halo}\ h^2\gsim 0.01-0.05$, where
$h$ is the reduced Hubble constant. 
On the opposite side, observations at large scales,
$(6-20)\ h^{-1} $ Mpc, have provided 
estimates of $\Omega_{CDM} h^2\approx 0.1-0.6$ 
\cite{freedman}, but values as low as 
$\Omega_{CDM} h^2\approx 0.02$ have also been quoted \cite{kaiser}.
Taking up-to-date limits on $h$, the 
baryon density from nucleosynthesis and overall matter-balance  
analysis one is able to obtain a favoured range,
$0.01\lsim \Omega_{CDM} h^2 \lsim 0.3$ (at $\sim 2\sigma$ CL)
 \cite{sadoulet,primack00}.
Note that  conservative 
lower limits in the small and large scales are 
of the same order of magnitude.

In this work, the expected neutralino cosmological 
relic density has been computed according to well 
known techniques (see \cite{review}). 
In principle, from its general behaviour  
$\Omega_{\chi} h^2\propto 1/ \langle \sigma_{ann}\rangle$,
where $\sigma_{ann}$ is the cross section for annihilation of 
neutralinos,
it is expected that such high neutralino-proton cross 
sections as those presented above  will then correspond to 
relatively low relic neutralino densities.
However, our results show that for some of the 
largest  cross-sections, with e.g. $M_I\approx 10^{12-11}$ GeV,
the value of the  relic density is  
still inside the conservative ranges we considered above.
As a function of the intermediate scale
 our results  show a steady 
increase in the value of the relic density 
when we move  from  
$M_I\approx 10^{11}$ GeV down to $M_I\approx 10^{16}$ GeV.
In this respect, the main conclusion to be drawn from our results
is that it should be always feasible, for large areas of supersymmetric
parameters, to find an ``intermediate'' initial scale $M_I$
which represents a compromise between a high neutralino cross section
and an adequate relic neutralino density. 

We expect that neutralino 
coannhilitations\footnote{The computation 
of the relic density carried out here,
following \protect\cite{review}, contains only a partial treatment 
of neutralino coannhilation channels.}
do not play an important role here 
since the mass differences among the LSP and the NLSP are not
too small in general. 
For example, for the relative mass difference among 
the two lightest neutralinos, we have found 
$\Delta m_{\tilde\chi^0_2 \tilde\chi^0_1} /
m_{\tilde\chi^0_1}\gsim 0.2$ in most of the interesting regions, 
being typically  much higher.

\section{Conclusions}

In this paper we have analyzed the relevant implications for 
dark matter analyses of the
possible existence of an initial scale $M_I$ 
smaller than $M_{GUT}$ to implement the boundary conditions. 

We have noted that the neutralino-nucleon cross section in the MSSM
is quite sensitive to the value of the initial scale for the running
of the soft breaking terms. The smaller the scale is, the
larger the cross section becomes. In particular, by taking
$M_I\approx 10^{10-12}$ GeV rather than $M_{GUT}\approx 10^{16}$ GeV
for the initial scale, which is a more sensible choice e.g. in the 
context of some superstring models, we find
that the cross section increases substantially being compatible
with the sensitivity of current dark matter experiments 
$\sigma_{\tilde\chi_1^0-p}\approx 10^{-6}$--$10^{-5}$ pb, 
for $\tan\beta\gsim 3$.
For larger values of the initial scale, as e.g. $M_I=10^{14}$ GeV,
the compatibility is obtained for $\tan\beta\gsim 10$.
Let us remark that these results have been obtained assuming
non-universal gauge couplings at $M_I$, as discussed in Section~2.
They should be compared with those of the MSSM with initial scale
$M_{GUT}$, where $\tan\beta\gsim 20$ is needed.

We have also discussed the corresponding relic neutralino 
densities and checked that they are of 
the right order of magnitude in large areas of the parameter space
for the neutralino 
being a CDM candidate.

The above computations have been carried out for the case of
universal soft terms. This is not only the most simple
possibility in the framework of the MSSM, but also 
is allowed in the context of superstring models. This is e.g.
the case of the dilaton-dominated supersymmetry breaking scenarios
or weakly and strongly coupled heterotic models with one K\"ahler modulus.
In this sense the analysis of neutralino-nucleon
cross sections of those models 
is included in our computation.

Obviously, non universality of the soft terms
will introduce more flexibility in the computation,
in particular in the value of $\mu^2$,
in order to obtain regions in the parameter space giving rise to  
cross sections compatible with the sensitivity
of current detectors.




\bigskip

\noindent {\bf Acknowledgments}

\noindent We thank P. Belli for interesting discussions about DAMA experiment.
We also thank G. Jungman for providing us with the relic-density 
code based on \cite{review}.

E. Gabrielli acknowledges the financial support
of the TMR network project ``Physics beyond the standard model'',
FMRX-CT96-0090. S. Khalil acknowledges the 
financial support of a Spanish Ministerio de Educaci\'on y Cultura 
research grant. 
The work of C. Mu\~noz has been supported 
in part by the CICYT, under contract AEN97-1678-E, and
the European Union, under contract ERBFMRX CT96 0090. 
The work of E. Torrente-Lujan was supported by a 
DGICYT grant AEN97-1678-E.


\newpage

\begin{figure}[htb]
\begin{center}
\begin{tabular}{c}
\epsfig{file= 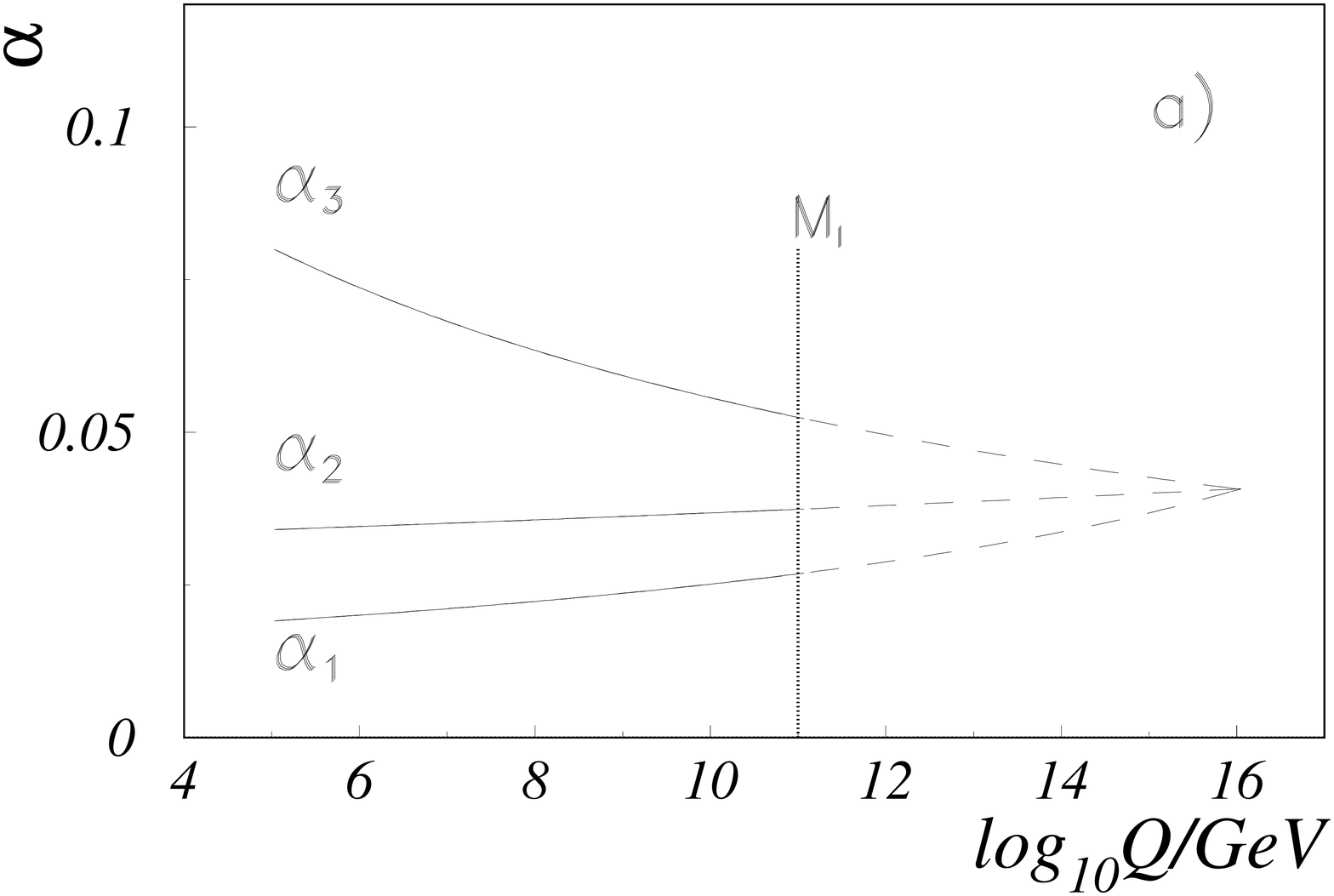, width=10cm, height=7.0cm}\\
\epsfig{file= 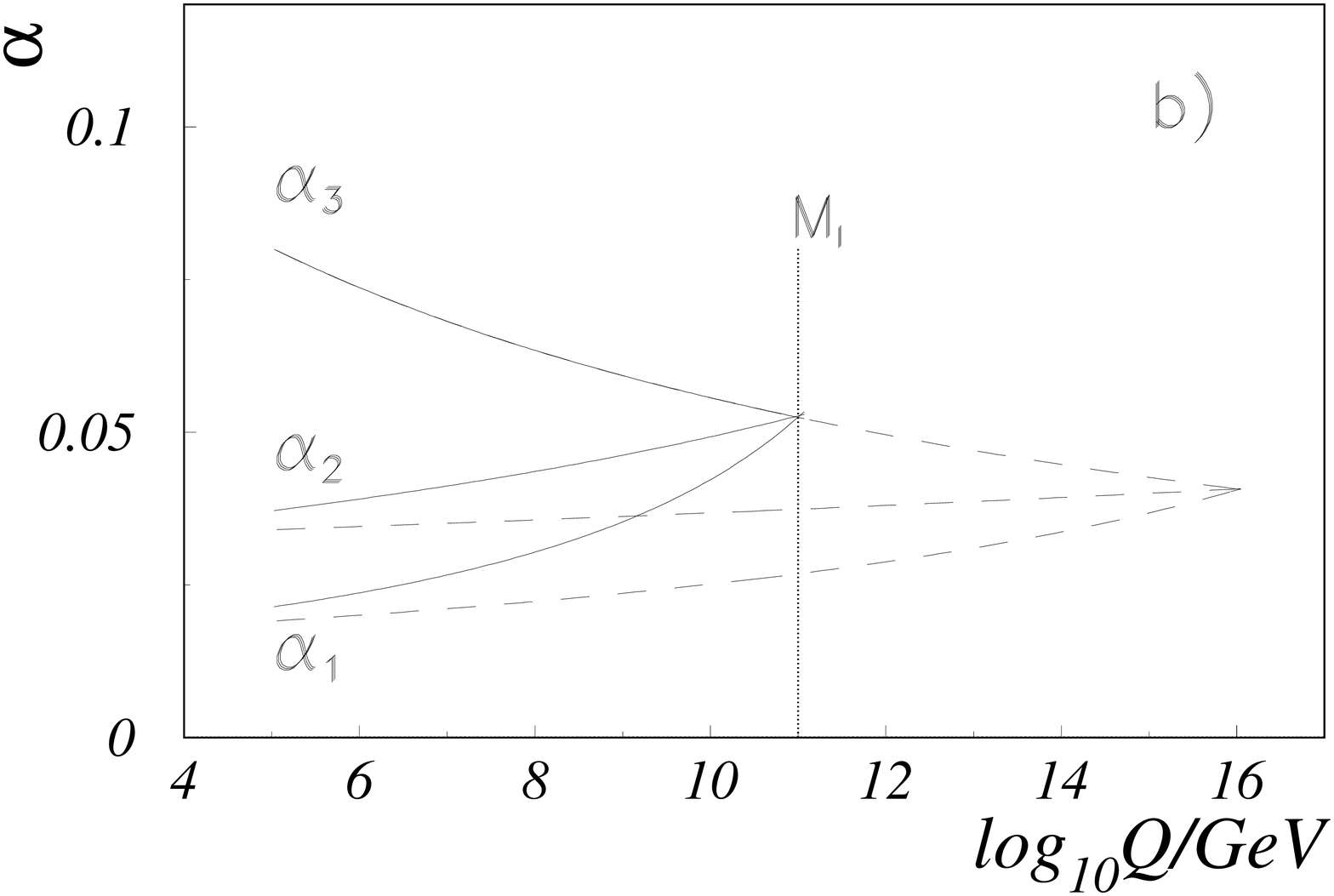,width=10cm,height=7.0cm}
\end{tabular}
\end{center} 
\caption{Running of the gauge couplings with energy 
assuming non universality (a) and
universality (b) of 
couplings at the initial scale 
$M_I=10^{11}$ GeV. Dashed lines indicate the usual running of the
MSSM couplings.} 
\end{figure}

\begin{figure}[htb]
\begin{center}
\begin{tabular}{c}
\epsfig{file= 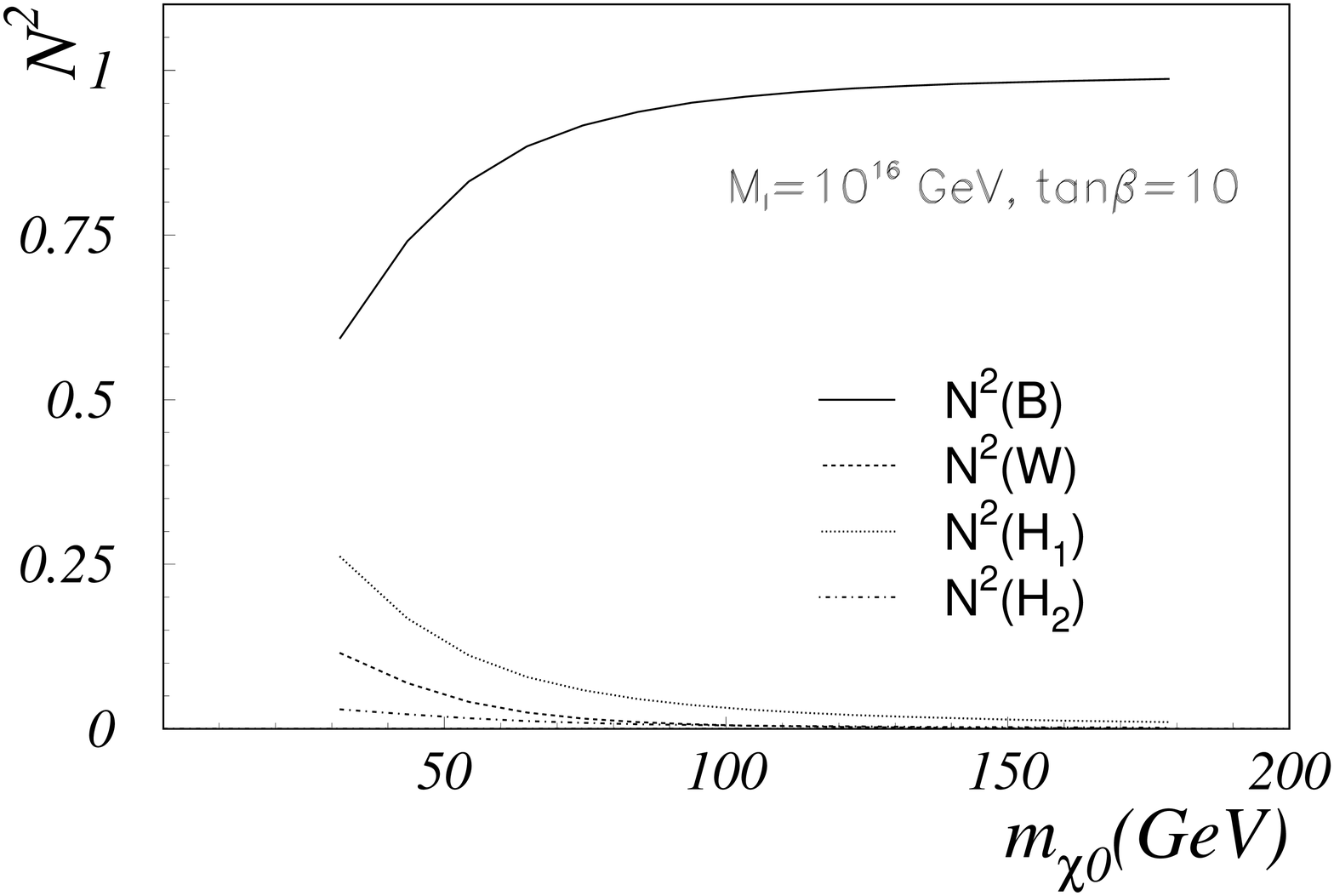,width=10cm,height=7.0cm}\\
\epsfig{file= 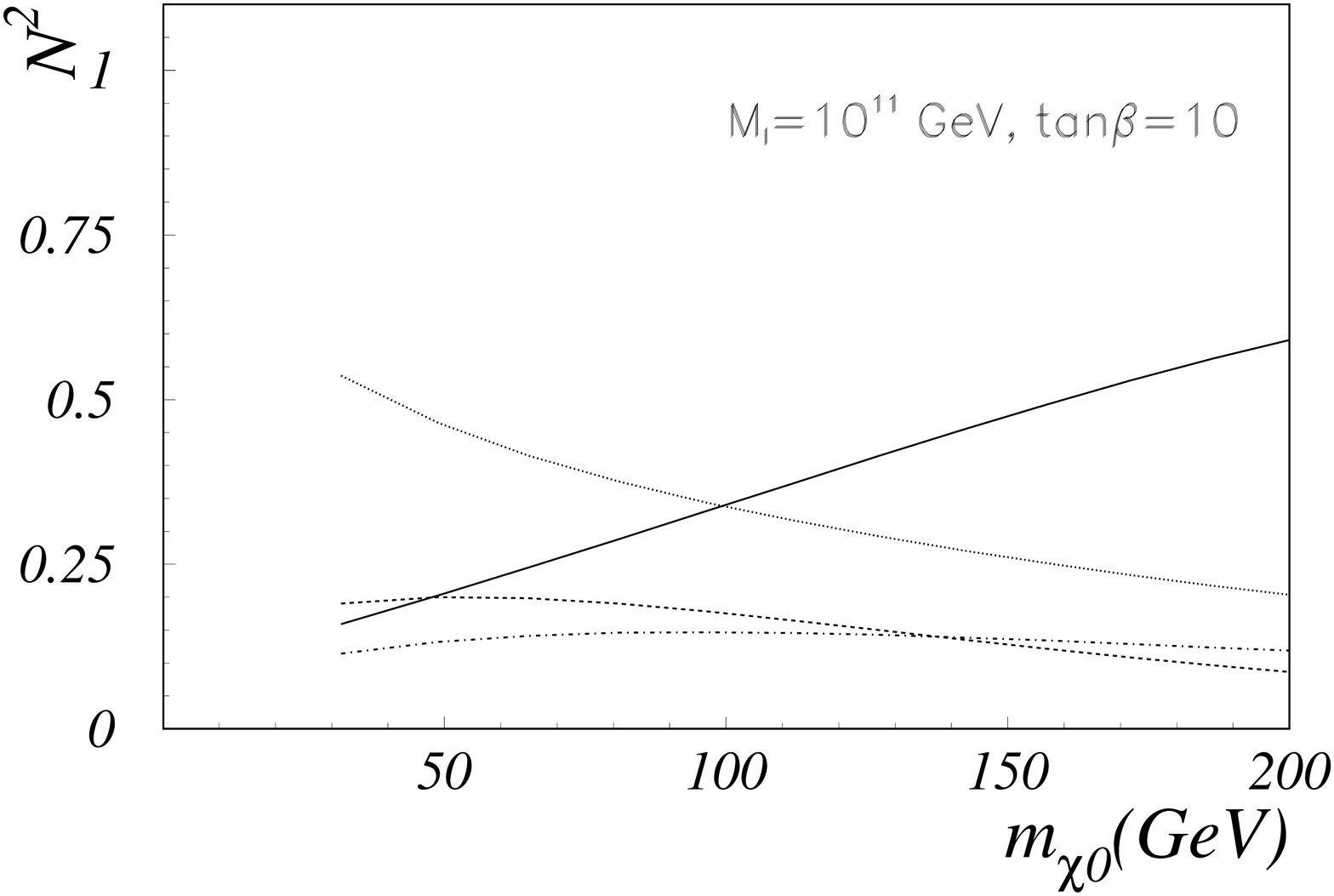, width=10cm, height=7.0cm}
\end{tabular}
\end{center} 
\caption{Gaugino-Higgsino components-squared $N_{1i}^2$ 
of the lightest neutralino as a function of its mass, for the case
$\tan\beta=10$, and 
for two different values of the initial scale, $M_I=10^{16}$ GeV and 
$M_I=10^{11}$ GeV.} 
\end{figure}

\begin{figure}[htb]
\begin{center}
\begin{tabular}{c}
\epsfig{file= 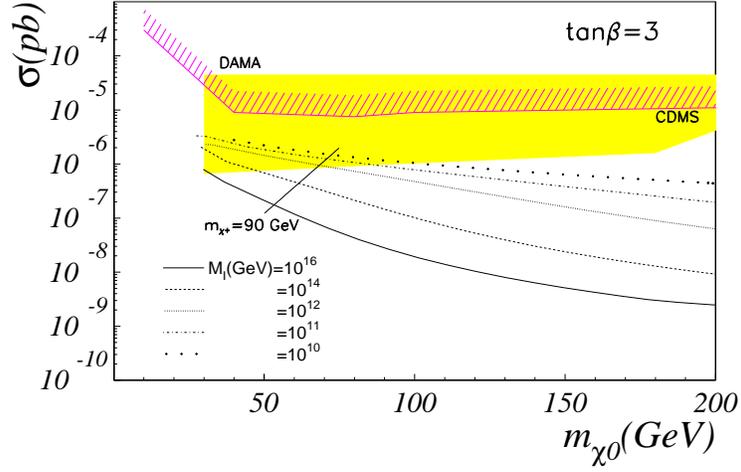, width=10cm, height=7.0cm}\\[-0.5cm]
\epsfig{file= 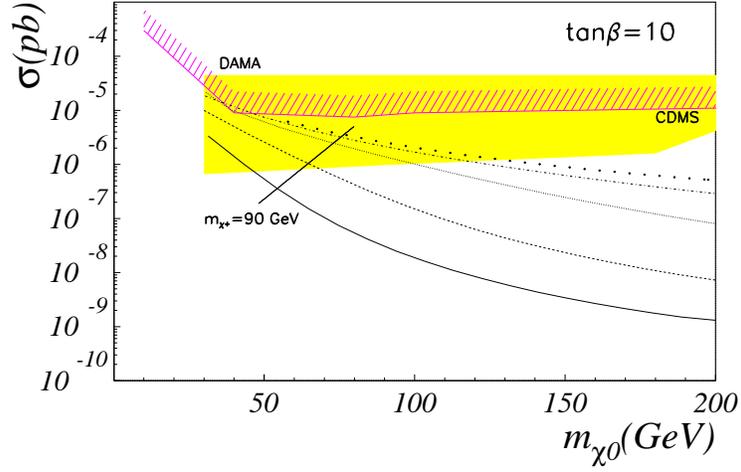,width=10cm,height=7.0cm}\\[-0.5cm]
\epsfig{file= 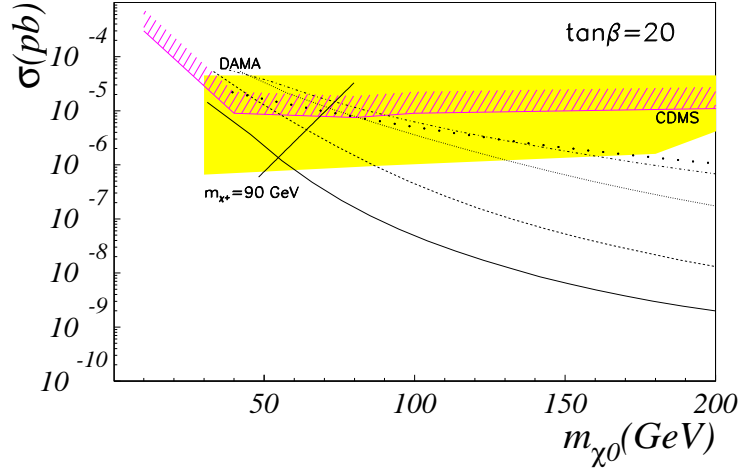, width=10cm, height=7.0cm}
\end{tabular}
\end{center} 
\vspace{-1.0cm}
\caption{Neutralino-proton cross section as a function of the
neutralino mass for several possible values of the initial scale
$M_I$, and for different values of $\tan\beta$. 
Current experimental limits, DAMA and CDMS, are shown. 
The region on the left of the 
line denoted by $m_{\tilde\chi_1^{\pm}}=90$ GeV is excluded because of the
experimental lower bound on the lightest chargino mass.} 
\end{figure}
\begin{figure}[htb]
\begin{center}
\epsfig{file= 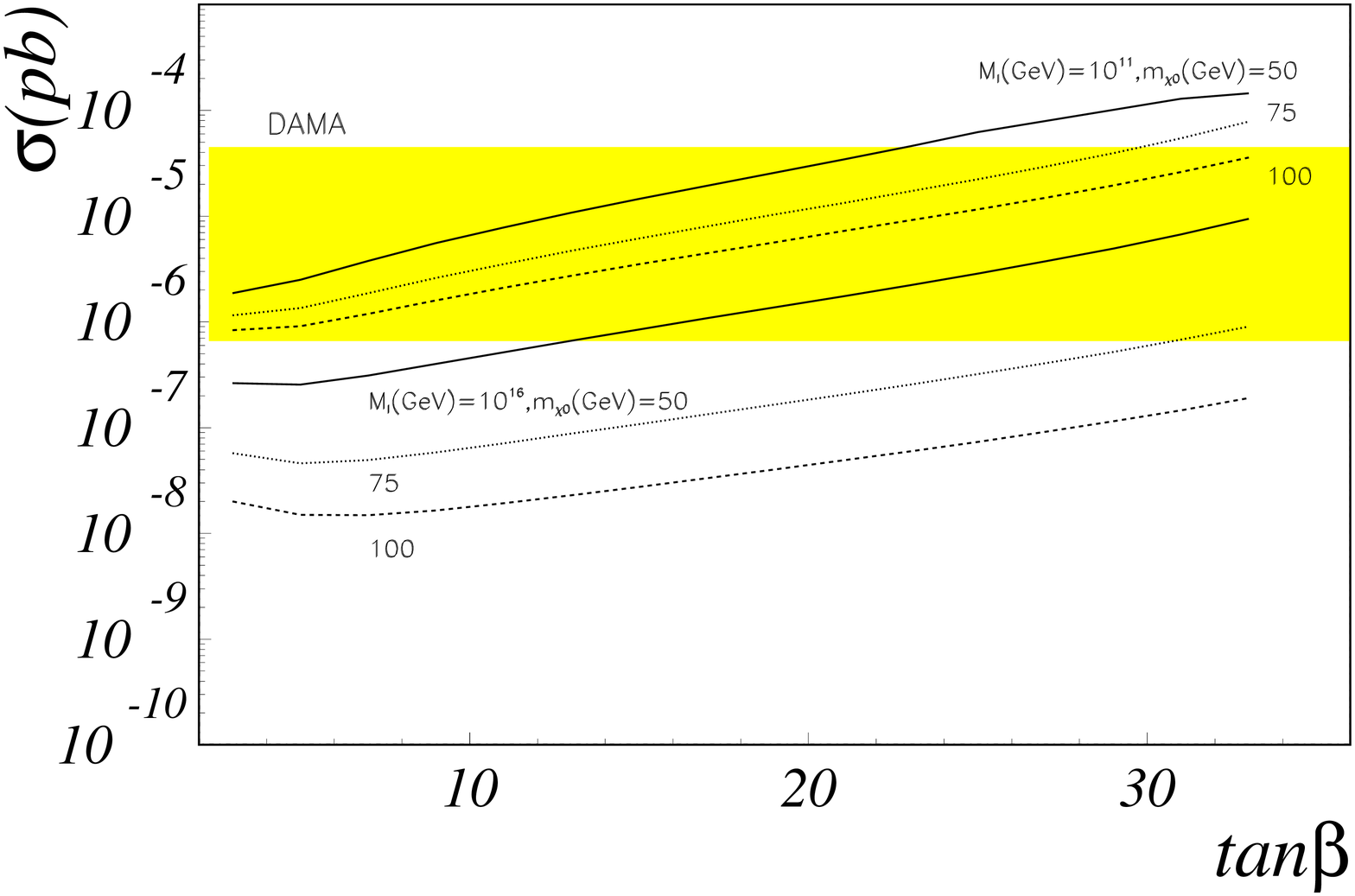, width=11cm, height=9cm}\\
\end{center} 
\caption{Neutralino-proton cross section as a function
of $\tan\beta$ for two possible values of the initial scale,
$M_I=10^{16}$ GeV, $M_I=10^{11}$ GeV, 
and for different values of the neutralino mass, namely
50, 75 and 100 GeV. DAMA limits are 
also shown.} 
\end{figure}

\begin{figure}[htb]
\begin{center}
\epsfig{file= 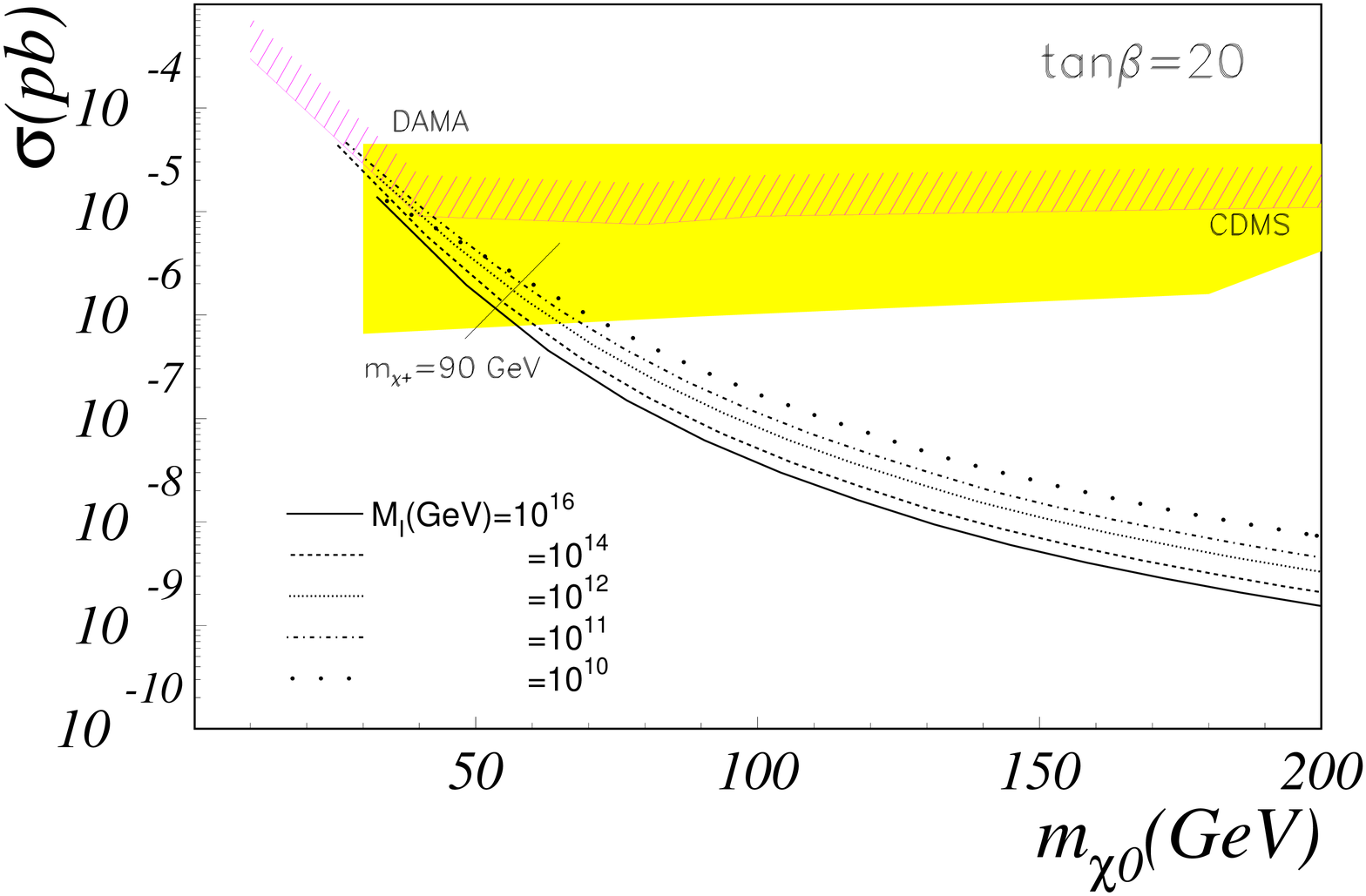, width=11cm, height=9cm}\\
\end{center} 
\caption{The same as in Fig.~3 but only for $\tan\beta =20$ and
with gauge coupling unification at $M_I$.} 
\end{figure}

\end{document}